# Non-equilibrium self-assembly of spin-wave solitons in FePt nanoparticles

D. Turenne[1], A. Yaroslavtsev[1,2], X. Wang[1], V. Unikandanuni[3], I. Vaskivskyi[4], M. Schneider[5], E. Jal[6], R. Carley[2], G. Mercurio[2], R. Gort[2], N. Agarwal[2], B. Van Kuiken[2], L. Mercadier[2], J. Schlappa[2], L. Le Guyader[2], N. Gerasimova[2], M. Teichmann[2], D. Lomidze[2], A. Castoldi[7,8], D. Potorochin[2,9,10,11], D. Mukkattukavil[1], J. Brock[12], N. Z. Hagström[3], A. H. Reid[13], X. Shen[13], X. J. Wang[13], P. Maldonado[1], Y. Kvashnin[1], K. Carva[14], J. Wang[15], Y. K. Takahashi[15], E. E. Fullerton[12], S. Eisebitt[5,16], P. M. Oppeneer[1], S. Molodtsov[2,10,11], A. Scherz[2], S. Bonetti[3,17], E. Iacocca[18,19], H. A. Dürr[1]*

1. Department of Physics and Astronomy, Uppsala University, 751 20 Uppsala, Sweden.
2. European XFEL GmbH, Holzkoppel 4, 22869 Schenefeld, Germany.
3. Department of Physics, Stockholm University, 106 91 Stockholm, Sweden.
4. Complex Matter Department, Jozef Stefan Institute, Ljubljana, Slovenia.
5. Max-Born-Institut, Berlin, Germany.
6. Sorbonne Université, CNRS, Laboratoire de Chimie Physique – Matière et Rayonnement, 75005 Paris, France.
7. Dipartimento di Elettronica, Informazione e Bioingegneria, Politecnico di Milano, Milano, Italy
8. Istituto Nazionale di Fisica Nucleare, Sezione di Milano, Milano, Italy
9. Deutsches Elektronen Synchrotron, 22607 Hamburg, Germany
10. Institute of Experimental Physics, Technische Universität Bergakademie Freiberg, 09599 Freiberg, Germany.
11. ITMO University, Kronverkskiy pr. 49, 197101 St. Petersburg, Russia.
12. Center for Memory and Recording Research, University of California San Diego, 9500 Gilman Drive, La Jolla, CA 92093-0401, USA
13. SLAC National Accelerator Laboratory, 2575 Sand Hill Road, Menlo Park, CA 94025, USA
14. Faculty of Mathematics and Physics, Department of Condensed Matter Physics, Charles University, Ke Karlovu 5, 121 16 Prague, Czech Republic.
15. Magnet Materials Unit, National Institute for Material Science, Tsukuba 305-0047, Japan.
16. Institut für Optik und Atomare Physik, Technische Universität Berlin, Berlin, Germany.
17. Department of Molecular Sciences and Nanosystems, Ca' Foscari University of Venice, 30172 Venice, Italy.
18. Department of Mathematics, Physics, and Electrical Engineering, Northumbria University, Newcastle upon Type, NE1 8ST, United Kingdom.
19. Center for Magnetism and Magnetic Materials, University of Colorado Colorado Springs, Colorado Springs, CO 80918, USA.

*Corresponding author. E-mail: hermann.durr@physics.uu.se

**Abstract**

Magnetic nanoparticles such as FePt in the $L1_0$-phase are the bedrock of our current data storage technology. As the grains become smaller to keep up with technological demands, the superparamagnetic limit calls for materials with higher magneto-crystalline anisotropy. This in turn reduces the magnetic exchange length to just a few nanometers enabling magnetic structures to be induced within the nanoparticles. Here we describe the existence of spin-wave solitons, dynamic localized bound states of spin-wave excitations, in FePt nanoparticles. We show with time-resolved X-ray diffraction and micromagnetic modeling that spin-wave solitons of sub-10 nm sizes form out of the demagnetized state following femtosecond laser excitation. The measured soliton spin-precession frequency of 0.1 THz positions this system as a platform to develop miniature devices capable of filling the THz gap.

**Teaser**

The smallest magnetic solitons known to date approaching their fundamental size limit have been observed in FePt nanoparticles.

**Introduction**

Spin waves are the fundamental excitations in magnetic systems. At low densities, they behave as independent quasiparticles that can mediate solid-state interactions such as superconducting pairing *(1)* or be used to transport information in technology *(2,3,4)*. At sufficiently high densities, spin waves can condense into solitons that derive their stability from non-linear spin precession *(5,6,7)*. Generation of spin-wave solitons requires a conservative environment *(8)*, where dissipation is matched by excitation, realized within spin-torque nano-contacts *(5,6,9)*. Non-equilibrium conditions via demagnetization with a femtosecond (fs) laser pulse provide an alternative generation mechanism *(10)* for topological spin textures, so-called skyrmions *(11,12,13)*. So far, the generated spin-wave solitons *(14,15,16)* and skyrmions *(11,12,13)* in materials with perpendicular magnetic anisotropy through femtosecond excitations are too large (several 100 nm) to be attractive for applications.

Ferromagnetic FePt nanoparticles are natural candidates for supporting spin-wave solitons *(8)* of the ultimate smallest size. The fundamental size limit is given by the so-called exchange length that in FePt is between 1-5 nm *(17)* and is thus significantly smaller than typical magnetic nanoparticle sizes (see Fig. 1). The exchange length describes the lengthscale on which a deviation from a homogeneous magnetic order can occur. It is determined by the competition of the magnetic exchange interaction, that aligns adjacent spins parallel (ferromagnetic) to one another, and the magneto-crystalline anisotropy, that in FePt favors atomic spins oriented along the so-called easy direction of magnetization (for FePt along the cylinder axis in Fig. 1). The magneto-crystalline anisotropy in FePt is extremely large *(17)* leading to small values of the exchange length and, thus, to domain-wall widths on the order of only a few atomic spacings, as schematically shown in the top inset of Fig. 1. In equilibrium, magneto-statics usually favors a single-domain magnetic order in nanoparticles that minimizes both exchange and anisotropy energies *(18)*. However, dynamic spin-wave solitons can theoretically exist in nanoparticles, as schematically shown in Fig. 1 for an edge soliton, i.e. one that is pinned to the nanoparticle's physical boundary.

Here we show that sub-10 nm spin-wave solitons are self-assembled in FePt nanoparticles following femtosecond laser excitation. Micromagnetic calculations identify the huge FePt magneto-crystalline anisotropy and pinning at the nanoparticle's boundary as key ingredients for soliton formation. The resulting characteristic soliton dynamics frequencies approach the THz regime. They are experimentally verified with time-domain X-ray scattering experiments via the strong FePt magneto-elastic coupling. These results establish a new nanoscale platform for exploring spin-wave solitons with only a few nanometers in size approaching the theoretical limit of the exchange length that has been elusive to date. This platform also opens the door to dramatically miniaturised information processing *(2,3,4)* and possibly bio-inspired computing applications *(19)*.

**Results**

We generate spin-wave solitons by taking advantage of the approach demonstrated in ref. *(10)* where a randomized spin distribution was induced by an ultrafast quench of the magnetic order after absorption of a fs laser pulse. This non-equilibrium demagnetized state is characterized by large-angle spin fluctuations via excited spin waves. Spin-wave solitons form by localization of long-wavelength spin waves that maintain the total energy of the system at short timescales *(20)*. This process is approximately captured by micromagnetic simulations for the small exchange lengths in FePt nanoparticles (see materials and methods). Figure 2 displays the resulting spin-wave soliton dynamics in a cylindrical FePt nanoparticle of 22.5 nm width and 8 nm height that is among the sizes commonly found in our samples (see also the supplementary Movie S1 of the soliton motion). The dynamics is characterized by precession of the in-plane magnetization, depicted in the soliton's perimeter (white region) in Figs. 2A, B. In addition, the spin-wave soliton is quickly attracted to the physical boundary *(21)* and experience both changes in their size (breathing or perimeter modes *(22)*) and translation along the nanoparticle's edge. These motions are apparent from the two snapshots displayed in Figs. 2A, B (additional snapshots are shown in Fig. S4). The characteristic frequencies involved in the in-plane precession and spin-wave soliton motion are shown in Figs. 2C and D, respectively. The in-plane precession is characterized by a sharp frequency peak centered around 0.05 THz. The spin-wave soliton motion in Fig. 2D contains two main spectral components: a broad frequency band around 0.10 THz that originates from the spin-wave soliton breathing and a low-frequency contribution <0.02 THz related to the soliton motion around the nanoparticle's boundary (see supplementary Movie S1).

To date, spin-wave solitons have been detected in extended magnetic thin films by directly imaging the reversed magnetization at the soliton core using X-rays *(14,15,16)*. In our case the much smaller soliton size (see Fig. 2) implies that this is below the resolution limits of typical magnetic X-ray imaging techniques *(23)*. We, therefore, resort to scattering techniques to probe the characteristic magnetization precession and spin-wave soliton breathing frequencies shown in Figs. 2C, D. The strong magneto-elastic coupling in FePt *(24,25)* provides a convenient means to achieve this goal. Typically, the magneto-elastic force acting on lattice atoms is directly related to the spatial gradient of the magneto-elastic energy change (see materials and methods). This implies that variations of the magnetization over very short distances can generate large displacements of lattice atoms as illustrated in the top inset of Fig. 1. The oscillatory nature of the magneto-elastic forces will then drive acoustic lattice waves that propagate throughout the FePt nanoparticles with the speed of sound (4.6 nm/ps for the longitudinal acoustic, LA, mode in FePt). The localized nature of the spin-wave-soliton-induced magneto-elastic forces causes the emitted acoustic waves to be coherent, i.e. the atoms vibrate with a fixed phase relationship. This situation is similar to what has been observed for acoustic strain waves generated at surfaces and interfaces of thin films *(26,27)*.

Figure 3 shows the time-domain measurement of the emitted coherent acoustic phonons in FePt nanoparticles. Measurements were performed at the SCS instrument of the European XFEL facility (see materials and methods). 30 fs X-ray pulses of 2500 eV photon energy were scattered as shown schematically in Fig. 3A (marked in blue) with the transferred wavevector, $q$, defined as indicated on the 2D detector. The FePt sample was heated by a 30 fs optical laser pulse (marked in red) intense enough to completely quench the FePt ferromagnetic order. The $q$-dependent scattering signal in Figs. 3B, C is dominated by an initial intensity drop caused by the laser-induced changes of the nanoparticle volume *(24)*. The size of this drop is consistent with a 1.4% lattice expansion at our employed pump fluence (see materials and methods).

Pronounced intensity oscillations are observed at times following the initial intensity drop in Figs. 3B, C. These oscillations correspond to coherent phonons composing lattice strain waves as observed previously for thin films *(26,27)*. The oscillation period displays characteristic variations with $q$ that are more clearly visualized in Fourier space. The time-frequency Fourier transform has the functional form $Ae^{i\varphi_0}$ where the determined frequency amplitude, $A$, is shown in Fig. 4A and the phase, $\varphi_0$, in Fig. 4B.

A feature with a linear dispersion seen in Fig. 4A at high frequencies and large $q$ is identified as propagating longitudinal acoustic (LA) phonons. The white line in Fig. 4A shows the calculated LA mode dispersion (see Fig. S7). In analogy to refs. *(26,27),* such phonons are excited as strain waves at the nanoparticle boundary and essentially are responsible for expanding the nanoparticle's volume. However, the most intense mode observed at 0.1 THz has virtually no group velocity.

To clarify the origin of the mode at 0.1 THz, we study the calculated scattering for the spin-wave soliton modes shown in Fig. 2. Details of the calculations are given in the materials and methods section. In brief, we use eqs. (1) and (2) to obtain the magneto-elastic lattice displacements, **u**, throughout the nanoparticle at each time step of the magnetization dynamics simulation (see Movie S1). The displacements, **u**, are the only input for scattering calculations using eq. (6). We azimuthally average the calculated scattering results to mimic the experimental conditions shown in Fig. 3A. Finally, we display in Fig. 4C the frequency amplitudes vs. $q$ obtained after time-frequency Fourier transform.

Comparison of Figs. 4A and 4C allows us to identify the fingerprints of spin-wave solitons in the experimental scattering data. While the high-frequency mode can be assigned to LA strain waves as described above, the frequency bands between 0.2-0.3 THz and especially the dominant mode at 0.1 THz are well reproduced by scattering from spin-wave solitons. The $q$-dependence of scattering amplitudes and phases for LA and spin-wave soliton modes are compared in Fig. 5 and will be discussed in the following section.

**Discussion**

The modes observed in Fig. 4 with frequencies of 0.2-0.3 THz and especially the even more intense feature at 0.10 THz do not agree with the expected LA-mode dispersion relation (white line). We can rule out that these modes are associated with TA modes (shown in Fig. S7) as the transverse polarization cannot be detected in our experimental geometry *(26, 27)*. In addition, other optical lattice modes have very different frequencies in FePt *(24)* outside the range shown in Fig. 4. Also, the coupling of phonon modes to spin waves can be ruled out since the lowest-energy spin-wave mode in FePt is energetically located above 0.69 THz (see Fig. S7). We note that ferromagnetic resonance (FMR) modes observed experimentally between 0.24-0.28 THz *(28)* cannot magneto-elastically couple to phonons as the oscillation amplitude of these FMR modes is nearly homogeneous across the nanoparticle.

Figures 5A-D display slices along the $q$-axis through the scattering amplitudes and phases of Figs. 4A, B at the selected frequencies 0.50 and 0.10 THz. At 0.50 THz (Figs. 5 A, B), the scattering amplitude in Fig. 5A is characterized by a single peak at $q = 0.57$ nm$^{-1}$ which agrees well with that expected from the theoretical LA-mode dispersion (Fig. 4A). The observed full width at half maximum, $\Delta q = 0.3$ nm$^{-1}$ corresponds to a frequency broadening of $\Delta \nu = 0.2$ THz which implies that the LA strain waves are heavily damped as indeed observed in the time-domain measurements of Fig. 3 B, C. In the simple picture of a driven harmonic oscillator the LA-mode phase should vary from 0 to $\pi$ when the driving frequency is swept across the LA resonance frequency. A significant part of this phase characteristics is indeed observed in Fig. 5B. At resonance the phase is close to zero while at low $q$-values which corresponds to frequencies below resonance we observe a phase of $-\pi/2$. At higher $q$-values the phase indeed starts to approach $\pi/2$ although the measured $q$-range is insufficient to actually reach this value.

The scattering characteristic from spin-wave solitons is markedly different. Figure 5C shows that the soliton scattering amplitude displays a two-peak structure. This is reproduced by the model shown in Fig. 5E. The model also allows us to assess the origin of these features. In particular, the dip in scattering amplitude observed at $q = 0.56$ nm$^{-1}$ is caused by scattering from the selected nanoparticle size, i.e. smaller (larger) nanoparticles exhibit the dip at larger (smaller) $q$-values. However, the relative intensity of the two peaks in the scattering amplitude observed at $q = 0.29$ nm$^{-1}$ and $q = 0.79$ nm$^{-1}$ is influenced by the soliton size. Solitons of smaller (larger) size will relatively scatter more (less) at larger $q$-values. The good agreement between measured (Fig. 5C) and calculated (Fig. 5E) amplitude q-dependence of the soliton scattering allow us to conclude that solitons of ~8 nm in size and a 0.05 THz spin-precession frequency are formed preferentially in nanoparticles of 22.5 nm diameter.

Solitons of a size slightly different to the ~8 nm shown in Fig. 5E should have frequencies that differ from the 0.1 THz magneto-elastic lattice motion driven by the 0.05 THz soliton spin precession. This may explain the broadening along the frequency axis observed around the 0.1 THz amplitude maximum in Fig. 4A. A closer inspection of linecuts along the $q$-axis for different frequencies display different $q$-dependencies. This is especially apparent at 0.13 THz where both the low-$q$ amplitude maximum and the amplitude dip occur at larger $q$-values compared to 0.10 THz. We modeled this behavior in micromagnetic simulations for different particle sizes and obtained good agreement with experiment for ~7 nm solitons (and a 0.65 THz spin precession frequency) in 19 nm nanoparticles. The amplitude scattering observed for frequencies below 0.1 THz indicates the existence of larger solitons in larger nanoparticles. We did not attempt to model this behavior in more detail due to the limited $q$-range of the data. However, the 0.1 THz amplitude maximum clearly indicates that the ~8 nm soliton size residing in 22.5 nm particles is the most abundant one in our samples.

The driving forces of the coherent lattice modes observed in Fig. 4 can be assessed by considering the phase relationship between these modes and the LA phonons. The phase, $\varphi_0$, of coherent oscillations in the time domain (Figs. 3B, C) describes the temporal offset with which the individual modes oscillate (see materials and methods). Fig. 4B shows the phase vs. *q* plot of the Fourier-transformed data from Fig. 3B. The peak positions visible in the amplitude plot of Fig. 4A are marked by the same lines also shown in Fig. 4B. The phase of the LA mode (for $q > 0.6$ nm$^{-1}$, i.e. where only the LA mode is clearly visible) is identical to that of the frequency band between 0.2-0.3 THz within the experimental error of ±0.3 radians.

However, the relative phase of the 0.10 THz mode is significantly different throughout the *q*-range (0.2-0.5 nm$^{-1}$) where it is visible. A detailed analysis of the 0.10 THz spin-wave soliton phase is more difficult than that of its scattering amplitude. This is largely due to the fact that phase changes observed along the frequency axis will extend far beyond the soliton resonance frequency in stark contrast to the phase originating from a relatively narrow amplitude peak. Consequently, a detailed modeling requires knowledge of the soliton size distribution in the sample. Rather than introducing additional fit parameters to account for the soliton size distribution we discuss here the salient features caused by it in the observed scattering phase shift. The dominant feature in the calculated soliton-scattering phase is a shift in Fig. 5F from $+\pi/2$ to $-\pi/2$ at a wavevector, $q = 0.6$ nm$^{-1}$, corresponding to the dip in scattering amplitude in Fig. 5E. It, thus, implies a zero crossing of the calculated scattering intensity. This feature is also clearly visible in the experimental phase values of Fig. 5D. However, experimentally the zero-crossing does occur with a phase offset that is close to $\pi/3$. We note that this phase value will to some degree be influenced by soliton resonances at neighboring frequencies and by a background due to the proximity to the LA resonance. The latter is likely also responsible to the deviation of the experimentally determined 0.10 THz-phase at *q*-values below 0.6 nm$^{-1}$.

Our data also allows to estimate the nucleation time of spin-wave solitons. The LA phonons are generated by the laser-induced lattice expansion that starts at the nanoparticle boundary. Their oscillatory lattice displacements composing the propagating strain wave essentially commence with the arrival of the pump laser pulse *(25,26)*. Also, the 0.2-0.3 THz modes start oscillating with the same phase, i.e. at the arrival time of the pump laser pulse. However, the 0.10 THz mode displays a phase lag (up to $\pi/3$) relative to the LA mode. If we assume that LA and 0.10 THz modes originate in similar regions of the nanoparticles, i.e. close to the nanoparticle boundaries, we can express the phase difference as a time delay which is given by the phase difference divided by the mode frequency as $\frac{\Delta\varphi_0}{2\pi\nu} \sim 1.7$ ps. Note that the sign between LA and 0.10 THz mode demonstrates that the latter starts oscillating ~1.7 ps later. We can, therefore, assign this value to the time it takes a spin-wave soliton to form out of the laser-demagnetized state.

Interestingly the calculations in Fig. 4C also show soliton related frequency features between 0.2-0.3 THz. These are due to magneto-elastic frequency mixing (for details see the Supplementary Materials section) between the frequency-doubled in-plane magnetization precession (Fig. 2C) and the spin-wave soliton breathing (Fig. 2D). It is tempting to assign them to the observed frequency band that is slightly blue-shifted with increasing wavevector (marked with orange lines in Fig. 4). However, the observed zero phase difference relative to the LA modes in Fig. 4B argues for a strain-wave-related origin of this frequency band. It is conceivable that strain waves are actually driving part of the soliton motion through magneto-elastically coupling back to the soliton magnetic dynamics. It has been shown that coherent elastic waves can drive spin precession modes in microstructures *(29)*. The microscopic origin is the effective magnetic field generated by magneto-elastic coupling *(29)*. In our case, the significantly larger strain wave amplitudes would lead to effective magnetic fields of several Tesla that could especially influence the $M_z$ magnetization dynamics responsible for the soliton breathing mode (see Fig. 2D). Such a mechanism could also explain the dispersion seen in the 0.2-0.3 THz mode (orange lines in Fig. 4) since the strain-wave propagation will naturally depend on the size of the nanoparticle that is selected by the transferred vector, *q*. However, the detailed modeling of this behavior is beyond the scope of the present paper.

Our results show conclusively that spin wave solitons form in FePt nanoparticles out of the demagnetized non-equilibrium state following heating with a fs optical laser pulse. We identify the coherent phonons generated by the spin-wave solitons' in-plane magnetization precession. The small magnetic exchange length of FePt determines the size of the spin-wave solitons of only several nanometers. This places the observed solitons squarely at the challenging boundary between the atomistic and continuous descriptions of magnetization dynamics *(30,31)*. We anticipate that our work will open up new theoretical and experimental efforts towards the understanding of magnetism at its intrinsic length and time scales, with implications for further scaling strategies in magnetic information storage and processing.

## Materials and Methods

<u>FePt sample growth and characterization.</u>
Single crystalline $L1_0$ FePt grains were grown epitaxially onto a single-crystal MgO(001) substrate by co-sputtering Fe, Pt and C *(32)*. This resulted in FePt nanoparticles of approximately cylindrical shape with heights of 8 nm and diameters in the range of 5–35 nm, with an average of 16 nm (see Fig. S1). The FePt nanoparticles form with *a* and *b* crystallographic directions, i.e. the $L1_0$ Fe and Pt planes, oriented parallel to the MgO surface. The space in-between the nanoparticles is filled with amorphous carbon. The film was covered by 50 nm of C acting as a heat sink for the pump-probe experiments. Following the sputtering process, the MgO substrate was chemically removed and the FePt-C films were floated onto copper wire mesh grids with 200-μm-wide openings.

We performed ultrafast electron diffraction from the FePt nanoparticles using the SLAC UED facility *(24)*. We deduce the FePt lattice expansion along the Fe and Pt atomic planes of the $L1_0$ structure (within the sample plane shown in Fig. 3A) as $\Delta a/a_0 = +1.4\% \pm 0.5\%$ for pump fluences up to 50 mJ/cm$^2$ (see Fig. S2) in agreement with Fig. 3B.

<u>Time-resolved X-ray diffraction experiments.</u>
The time-resolved tender X-ray diffraction experiments were performed at the Spectroscopy and Coherent Scattering (SCS) instrument of European XFEL at the photon energy 2500 eV. The soft X-ray monochromator grating set up in the second diffraction order provides an X-ray bandwidth around 400 meV at 2500 eV and suppresses the higher harmonics. The array of FePt samples were installed on a sample holder that could be moved in all three spatial directions relative to the beam. The X-ray beam was focused on the sample to a spot size of 80 μm using a Kirkpatrick-Baez mirror system. The X-ray fluence on the sample was approximately 0.5 mJ/cm$^2$. All measurements were performed at normal X-ray incidence.

The X-ray diffraction patterns were measured using the DSSC detector equipped with MiniSDD sensors *(33)*, at a distance of 184 cm from the sample. The DSSC detector has a 1024 × 1024 pixel matrix split into 16 sensors, 128 × 512 pixels each, grouped into 4 quadrants. The pixels of size 236 × 204 µm² are arranged in the sensors hexagonally. The matrix is covered by a thin Al filter to prevent any optical contamination of the detector image. During the data analysis the hexagonal pixel array was converted into squares leading to a negligible error for the count conserving transformation. A mask was applied to the measured patterns in order to exclude signals from "bad" pixels and residual stray light from upstream beamline elements. The incoming X-ray pulse energy was measured with the X-ray Gas Monitor (XGM) detector. This value was used for the normalization of diffraction patterns obtained for each X-ray shot.

The pump fs laser used was set to the fundamental wavelength of 800 nm. Laser and X-ray beams are combined in the laser in-coupling (LIN) chamber, approximately 1 m upstream from the sample. The spatial overlap between the X-ray and laser beams was verified by microscope camera images. The temporal overlap was verified in two stages. Coarse timing was done using the overlap of X-ray and laser signals measured by a photo diode connected to a fast oscilloscope. Fine timing was achieved by measuring the X-ray pump – laser probe reflectivity from a silicon nitride membrane, installed on the sample holder in the same plane as the FePt samples. The laser spot size on the sample was 170 µm and the pump fluence was 50 mJ/cm². The time resolution of the pump-probe experiment was 60 fs. Experiments were performed at 10 Hz repetition rate using laser pump – X-ray probe pulses as well as another X-ray pulse arriving approximately 70 µs earlier to probe the initial state of the sample.

The resulting detector images were background substracted, binned according to time delay and then normalized to the incoming X-ray fluence obtainged from the XGM. In order to reduce the data size and exploit the symmetries of the system, the scattering patterns were azimuthally integrated and the intensity as a function of transferred wavector, $q$ (see Fig. 3A), was obtained. The time dependent scattering pattern was normalized to the ground-state scattering at negative time delays, i.e. the laser-induced differences (with a constant offset of order unity) in the scattering patterns are shown throughout this paper.

We also performed X-ray scattering measurments at SCS with the X-ray energy in resonance with the Fe 2*p*-3*d* core-valence resonance at 708 eV. This allowed us to determine the amount of FePt demagnetization in analogy to refs. *(24,34)*.

Micromagnetic simulations.

The magnetization dynamics of isolated nanoparticles were simulated with the GPU package MuMax 3.9 *(35)*. We used magnetic parameters for FePt as measured in ref. *(28)*: saturation magnetization $M_s = 950$ kA/m, uniaxial anisotropy field $\mu_0 H_k = 8.9$ T leading to an energy density of $K_u = 4227.5$ kJ/m³ and Gilbert damping coefficient $\alpha = 0.1$. The used exchange constant of $A = 4.1$ pJ/m leads to an exchange length $l_{ex} = 3.1$ nm. We used micromagnetic cells of size 0.7 nm × 0.7 nm × 0.5 nm which were found to accurately resolve the dynamics by use of an adaptive Runge-Kutta 45 stepper limited to an upper time step of 1 ps. The simulations presented here pertain to a circular nanoparticle of 22.5 nm in diameter and 8 nm in thickness, resulting in a simulation domain of 32 × 32 × 16 = 16384 cells. This is close to the average nanoparticle size as determined from analysis of a TEM image (see Fig. S1). Use of the edge smoothing option in MuMax did not qualitatively affect the results.

Two distinct simulations for a single nanoparticle were performed. First, the remagnetization after ultrafast quenching was modeled as the evolution of the magnetization from a spatially uniform random distribution. This is a crude approximation for the first few to 10 ps, which is better described by atomistic spin dynamics, but yields qualitatively accurate results when the short-wavelength features are relaxed *(10)*. The time step in these simulations is typically on the order of 10 attoseconds. We observe the nucleation of solitons akin to magnon localization and coalescence for extended magnetic films with perpendicular magnetic anisotropy *(10,13)*. A well-defined edge soliton is observed at 80 ps of simulation time. After ~150 ps, the nanoparticle relaxes into a homogeneous magnetization.

To analyze the soliton dynamics, we perform a second set of simulations where the dissipation is disabled by setting the damping parameter $\alpha = 0$. We use the soliton relaxed at 80 ps as an initial condition and let the simulation run for 100 ps with a sampling of 50 fs. The goal of this conservative simulation is to numerically extract the soliton modes (breathing, motion, and perimeter in-plane magnetization precession) by Fourier analysis. With the used sampling and simulation time, we obtain a spectral resolution of 10 GHz and an upper frequency of 10 THz.

Magneto-elastic coupling and lattice dynamics calculations.
In order to evaluate the response of the FePt atomic structure to the presence of spin-wave soliton we performed magneto-elastic calculations using the results of micromagnetic simulations as an input. The spatially localized spin-wave soliton magnetization dynamics causes a strong magneto-elastic force, $\mathbf{f}_{mel}$, acting on the lattice-atom displacements, $\mathbf{u}$, via *(36)*

$$\rho \frac{\partial^2 \mathbf{u}}{\partial t^2} + \frac{2\rho}{\tau} \frac{\partial \mathbf{u}}{\partial t} = \nabla \boldsymbol{\sigma} + \mathbf{f}_{mel} \qquad (1)$$

Here $\rho$ is the mass density, $\tau$ is a damping time constant, $\boldsymbol{\sigma}$ is the stress tensor and $\nabla \boldsymbol{\sigma}$ is the elastic force per unit volume which defines the elastic properties of material. It is determined by the elastic stiffness constants given in Table 1 and by the elastic strains in the lattice *(36)*. We have derived the expression of the magneto-elastic force for the tetragonal lattice as *(37,38)*

$$\mathbf{f}_{mel} = \frac{1}{M_0^2} \begin{bmatrix} b_3 \frac{\partial M_x^2}{\partial x} \\ b_3 \frac{\partial M_y^2}{\partial y} \\ b_{22} \frac{\partial M_z^2}{\partial z} \end{bmatrix} + \frac{1}{2M_0^2} \begin{bmatrix} (b_3 + 2b_{21}) \frac{\partial M_z^2}{\partial x} \\ (b_3 + 2b_{21}) \frac{\partial M_z^2}{\partial y} \\ 0 \end{bmatrix} + \frac{1}{M_0^2} \begin{bmatrix} b_3' \frac{\partial}{\partial y}(M_x M_y) + b_4 \frac{\partial}{\partial z}(M_x M_z) \\ b_3' \frac{\partial}{\partial x}(M_y M_x) + b_4 \frac{\partial}{\partial z}(M_y M_z) \\ b_4 \frac{\partial}{\partial x}(M_z M_x) + b_4 \frac{\partial}{\partial y}(M_z M_y) \end{bmatrix} \qquad (2)$$

where $M_{x,y,z}$ are the components of the magnetization vector, $\mathbf{M}$, $M_0$ is its size and $b_{21}, b_{22}, b_3, b_3', b_4$ are tetragonal magneto-elastic coupling parameters. The density-functional theory based MAELAS program *(37)* was used to compute the magneto-elastic parameters, given in Table 1.

The equation of motion, eq. (1), was solved numerically in the 3-dimensional Cartesian grid (similar as in the micromagnetic simulation) using the standard second order "leapfrog" algorithm from the central differences. The dissipation term with $\tau = 5$ ps was included in eq. (1) to address, in the generalized form, the damping of magneto-elastically induced lattice vibrations via transmission through the nanoparticle boundary into the carbon matrix and other possible mechanisms. From the Fourier analysis of atomic displacements, the characteristic frequencies of lattice vibrations were obtained and are compared to the experimental results in Fig. 4. In general, the magneto-elastic coupling results in a doubling of the soliton large-angle precession mode frequencies, since the force eq. (2) contains the products of various magnetization components. In turn, the smaller soliton breathing mode amplitudes can be treated in a linearized way. More details are given in the Supplementary Materials section.

X-ray diffraction from coherent phonons in FePt nanoparticles.
The scattering intensity at a transferred wavevector **q** from a solid can be expressed as *(39)*

$$I(\mathbf{q}) = I_e | \sum_n f_n e^{i\mathbf{q}\cdot\mathbf{r}_n} |^2 \quad (3)$$

where $f_n$ are the atomic scattering factors for atom, $n$, and $\mathbf{r}_n$ are the atomic position vectors. Typically $I_e$ describes scattering from an individual electron *(39)*. However, in our experimental geometry $I_e$ also desrcibes the X-ray transmission through the sample *(40)*.

Eq. (3) can be used to estimate the change in scattering intensity upon lattice expansion following laser heating (see Fig. 3). The atomic scattering factors, $f_n$, are given by the tabulated optical constants *(41)* that scale inversely proprtional to the lattice unit cell volume, i.e. the atomic density. When the unit-cell volume increases due to laser-heating the scattering intensity changes inversely proprtional to it. We note that in our experimental geometry the incoming X-ray beam averages over the spatial coordinate perpendicular to the sample plane. As a consequence we only need to take the unit-cell expansion perpendicular to the X-ray incidence direction into account. Using the experimentally determined expansion of 1.4% (see pervious methods paragraph) we can explain the observed drop in the scattering intensity by 6.6% ± 2.1% (see Fig. 3C).

Eq. (3) is also the starting point to describe diffuse X-ray scattering from thermally and optically excited phonons *(26,27,39)*. Rewriting the absolute square in eq. (3) as

$$I(\mathbf{q}) = I_e \sum_{n,n'} f_n f_{n'} \, e^{i\mathbf{q}\cdot(\mathbf{r}_n^0 - \mathbf{r}_{n'}^0)} \, e^{i\mathbf{q}\cdot(\mathbf{u}_n - \mathbf{u}_{n'})} \quad (4)$$

the atomic displacements, $\mathbf{u}_n$, around the atomic positions at rest, $\mathbf{r}_n^0$, are then replaced by phonons of wavevector, **k**, and phonon branch, $s$, as *(39)*

$$\mathbf{u}_n = \operatorname{Re} \frac{1}{\sqrt{\mu}} \sum_{\mathbf{k},s} a_{\mathbf{k},s} \mathbf{e}_{\mathbf{k},s} e^{i\mathbf{k}\cdot\mathbf{r}_n^0 - i\omega_{\mathbf{k},s} t + i\varphi_{\mathbf{k},s}} \quad (5)$$

where $\mu$ is the atomic mass, $\omega_{\mathbf{k},s}$ is the phonon frequency and $\varphi_{\mathbf{k},s}$ a phase factor. $a_{\mathbf{k},s}$ and $\mathbf{e}_{\mathbf{k},s}$ describe phonon amplitude and polarization, respectively. It is important to note that while $\varphi_{\mathbf{k},s}$ in thermal equilibrium is random and averages to zero *(39)*, in our case $\varphi_{\mathbf{k},s}$ is the same for all phonons as long as they are generated by the same spatially localized force. This can either be the spin-wave solitons described in the main text or, coherent lattice strain waves due to lattice expansion starting at the nanoparticle boundary.

In order to evaluate eqs. (4) and (5) it is common to expand the term in eq. (4) containing the atomic displacements, **u**, as $e^{i\mathbf{q}\cdot(\mathbf{u}_n-\mathbf{u}_{n\prime})} = 1 + i\mathbf{q}\cdot\mathbf{u}_n - i\mathbf{q}\cdot\mathbf{u}_{n\prime} + O(u^2)$. In thermal diffuse scattering the linear terms average to zero and, therefore, the quadratic terms are used to describe the phonon contributions *(39)*. This also applies to time-resolved measurements of incoherently excited phonons *(42,43)*. Here we use the linear terms that give rise to scattering from coherent phonon wavepackets as demonstrated for thin films *(26,27)*. We arrive at the scattering from coherent phonons as

$$I_1(\mathbf{q}) \propto A(\mathbf{q}) \, \text{Im} \sum_n f_n \, e^{i\mathbf{q}\cdot\mathbf{r}_n^0} \, \mathbf{q}\cdot\mathbf{u}_n$$

$$= A(\mathbf{q}) \, \text{Im} \sum_n f_n \, e^{i\mathbf{q}\cdot\mathbf{r}_n^0} \text{Re} \frac{1}{\sqrt{\mu}} \sum_{\mathbf{k},s} a_{\mathbf{k},s} \mathbf{q}\cdot\mathbf{e}_{\mathbf{k},s} e^{i\mathbf{k}\cdot\mathbf{r}_n^0 - i\omega_{\mathbf{k},s}t + i\varphi_0} \quad (6)$$

where $A(\mathbf{q}) \propto \sum_n f_n e^{i\mathbf{q}\cdot\mathbf{r}_n^0}$ is the scattering amplitude from the atoms at rest. $A(\mathbf{q})$ is a real function for the cylindrical nanoparticles considered here. Eq. (6) represents the Fourier transform (n-summation) over a set of waves propagating in direction, **k**, with constant phase, $\varphi_0$. The term, $\mathbf{q}\cdot\mathbf{e}_{\mathbf{k},s}$, implies that phonons with a polarization vector parallel to the scattered wavevector are preferentially detected. It is important to reiterate that the phase term $e^{i\varphi_0}$ in eq. (6) is characteristic for the force that generates the coherent phonons. This is used in Fig. 4 to differentiate between phonons generated via strain waves at the nanoparticle boundary and phonons generated by spin-wave solitons.

We use eq. (6) to calculate the scattering pattern from spin-wave solitons. The spin-wave soliton magnetization dynamics (see Fig. 2 and Figs. S4A-F) generates a magneto-elastic force that acts on the lattice atoms (see Fig. S5A). The coresponding atomic displacements, $\mathbf{u}_n$, are calculated from eqs. (1) and (2). In these calculations we take $A(\mathbf{q})$ to be constant, i.e. without any **q**-dependence. This procedure reflects the normalization of the time-resolved X-ray diffraction measursments by the scattering yield before time zero, i.e. the FePt ground-state configuration. Figure 4C shows the calculated scattering amplitude of the spin-wave soliton mode. In particlular, the *q*-characteristics of the spin-wave soliton scattering near 0.10 THz closely resembles the experimental result in Fig. 4A.

Calculations of FePt phonons and spin waves.
The magnon dispersion of bulk FePt was calculated using a density-functional theory (DFT)-based approach. First, the DFT electronic structure of FePt was computed using the tight-binding linear muffin-tin method (TB-LMTO) within the atomic sphere approximation *(44)*. The DFT exchange-correlation potential was described by the local spin density approximation (LSDA) in the parametrization of Vosko *et al. (45)*. This approach has been used recently to study the atomic magnetic moments on Fe and Pt in FePt *(46)*. The magnon spectrum was subsequently computed by mapping the total energy on the Heisenberg model *(47)*. The effective pair exchange interactions $J_{ij}$ of the Heisenberg model that are required were computed using the Liechtenstein formula *(48)*.

The computed lowest-energy magnon dispersion of FePt is shown in Fig. S6. The magneto-crystalline anisotropy energy (MAE) leads to an upward shift of the spin-wave energy around the Γ point by the MAE (for one FePt unit) by ~0.69 THz *(49)*, in good agreement with other calculations *(50)* and inelastic neutron scattering measurements *(51)*.

We performed phonon-calculations following ref. *(24)*. This resulted in values of the speed of sound of 4.6 nm/ps for the LA and 2.6 nm/ps and 1.7 nm/ps for the TA phonon modes. Frequency dispersions of the phonon modes are shown in Fig. S7 together with the lowest-energy magnon mode.

**References**
1. T. Dahm, V. Hinkov, S. V. Borisenko, A. A. Kordyuk, V. B. Zabolotnyy, J. Fink, B. Büchner, D. J. Scalapino, W. Hanke, B. Keimer, Strength of the spin-fluctuation-mediated pairing interaction in a high-temperature superconductor. *Nat. Phys.* **5**, 217 (2009).
2. A. Chumak, V. Vasyuchka, A. Serga, B. Hillebrands, Magnon spintronics. *Nat. Phys.* **11**, 453–461 (2015).
3. S. Neusser, D. Grundler, Magnonics: Spin waves on the nanoscale. *Adv. Mater.* **21**, 2927 (2009).
4. B. Lenk, H. Ulrichs, F. Garbs, M. Münzenberg, The building blocks of magnonics. *Phys. Rep.* **507**, 107 (2011).
5. A. Slavin, V. Tiberkevich, Spin Wave Mode Excited by Spin-Polarized Current in a Magnetic Nanocontact is a Standing Self-Localized Wave Bullet. *Phys. Rev. Lett.* **95**, 237201 (2005).
6. S. Bonetti, V. Tiberkevich, G. Consolo, G. Finocchio, P. Muduli, F. Mancoff, A. Slavin, J. Åkerman, Experimental Evidence of Self-Localized and Propagating Spin Wave Modes in Obliquely Magnetized Current-Driven Nanocontacts. *Phys. Rev. Lett.* **105**, 217204 (2010).
7. M. A. Hoefer, T. J. Silva, M. W. Keller, Theory for a dissipative droplet soliton excited by a spin torque nanocontact. *Phys. Rev. B* **82**, 054432 (2010).
8. A. M. Kosevich, B. A. Ivanov, A. S. Kovalev, Magnetic Solitons. *Phys. Rep.* **194**, 117 (1990).
9. S. M. Mohseni, S. R. Sani, J. Persson, T. N. Anh Nguyen, S. Chung, Ye. Pogorylov, P. K. Muduli, E. Iaccoca, A. Eklund, R. K. Dumas, S. Bonetti, A. Deac, M. A. Hoefer, J. Åkermann, Spin Torque–Generated Magnetic Droplet Solitons. *Science* **339**, 1295 (2013).
10. E. Iacocca, T.-M. Liu, A. H. Reid, Z. Fu, S. Ruta, P. W. Granitzka, E. Jal, S. Bonetti, A. X. Gray, C. E. Graves, R. Kukreja, Z. Chen, D. J. Higley, T. Chase, L. Le Guyader, K. Hirsch, H. Ohldag, W. F. Schlotter, G. L. Dakovski, G. Coslovich, M. C. Hoffmann, S. Carron, A. Tsukamoto, M. Savoini, A. Kirilyuk, A. V. Kimel, Th. Rasing, J. Stöhr, R. F. L. Evans, T. Ostler, R. W. Chantrell, M. A. Hoefer, T. J. Silva, H. A. Dürr, Spin-current-mediated rapid magnon localisation and coalescence after ultrafast optical pumping of ferrimagnetic alloys, *Nat. Commun.* **10**:1756 (2019).
11. S.-G. Je, P. Vallobra, T. Srivastava, J.-C. Rojas-Sańchez, T. H. Pham, M. Hehn, G. Malinowski, C. Baraduc, S. Auffret, G. Gaudin, S. Mangin, H. Beá, O. Boulle, Creation of Magnetic Skyrmion Bubble Lattices by Ultrafast Laser in Ultrathin Films. *Nano Lett.* **18**, 7362 (2018).
12. M. Finazzi, M. Savoini, A. R. Khorsand, A. Tsukamoto, A. Itoh, L. Duo, A. Kirilyuk, Th. Rasing, M. Ezawa, Laser-Induced Magnetic Nanostructures with Tunable Topological Properties. *Phys. Rev. Lett.* **110**, 177205 (2013)
13. F. Büttner, B. Pfau, M. Böttcher, M. Schneider, G. Mercurio, C. M. Günther, P. Hessing, C. Klose, A. Wittmann, K. Gerlinger, L.-M. Kern, C. Strüber, C. von Korff Schmising, J. Fuchs, D. Engel, A. Churikova, S. Huang, D. Suzuki, I. Lemesh, M. Huang, L. Caretta, D. Weder, J. H. Gaida, M. Möller, T. R. Harvey, S. Zayko, K. Bagschik, R. Carley, L. Mercadier, J. Schlappa, A. Yaroslavtsev, L. Le Guyarder, N. Gerasimova, A. Scherz, C. Deiter, R. Gort, D. Hickin, J. Zhu, M. Turcato, D. Lomidze, F. Erdinger, A. Castoldi, S. Maffessanti, M. Porro, A. Samartsev, J. Sinova, C. Ropers, J. H. Mentink, B. Dupé, G. S.

D. Beach, S. Eisebitt, Observation of fluctuation-mediated picosecond nucleation of a topological phase. *Nat. Mater.* **20**, 30 (2021).
14. D. Backes, F. Macià, S. Bonetti, R. Kukreja, H. Ohldag, A. D. Kent, Direct Observation of a Localized Magnetic Soliton in a Spin-Transfer Nanocontact. *Phys. Rev. Lett.* **115**, 127205 (2015).
15. S. Bonetti, R. Kukreja, Z. Chen, F. Macia, J. M. Hernandez, A. Eklund, D. Backes, J. Frisch, J. Katine, G. Malm, S. Urazhdin, A. D. Kent, J. Stöhr, H. Ohldag, H. A. Dürr, Direct observation and imaging of a spin-wave soliton with p−like symmetry, *Nat. Commun.* **6**, 8889 (2015).
16. S. Chung, Q. T. Le, M. Ahlberg, A. A. Awad, M. Weigand, I. Bykova, R. Khymyn, M. Dvornik, H. Mazraati, A. Houshang, S. Jiang, T. N. A. Nguyen, E. Goering, G. Schütz, J. Gräfe, J. Åkerman, Direct Observation of Zhang-Li Torque Expansion of Magnetic Droplet Solitons. *Phys. Rev. Lett.* **120**, 217204 (2018).
17. K. Piao, D. Li, D. Wei, The role of short exchange length in the magnetization processes of $L1_0$-ordered FePt perpendicular media. *J. Magn. Magn. Mater.* **303**, e39 (2006).
18. A. Hubert, R. Schäfer, Magnetic domains: the analysis of magnetic microstructures. Springer (2009).
19. M. Romera, P. Talatchian, S. Tsunegi, F. A. Araujo, V. Cros, P. Bortolotti, J. Trastoy, K. Yakushiji, A. Fukushima, H. Kubota, J. Yuasa, M. Ernoult, D. Vodenicarevic, T. Hirtzlin, N. Locatelli, D. Querlioz, J. Grollier, Vowel recognition with four coupled spin-torque nano-oscillators. *Nature* **563**, 230 (2018).
20. B. Rumpf, A. C. Newell, Coherent Structures and Entropy in Constrained, Modulationally Unstable, Nonintegrable Systems. *Phys. Rev. Lett.* **87**, 054102 (2001).
21. E. Iacocca, R. K. Dumas, L. Bookman, M. M. Mohseni, S. Chung, M. A. Hoefer, J. Åkerman, Confined dissipative droplet solitons in spin-valve nanowires with perpendicular magnetic anisotropy. *Phys. Rev. Lett.* **112**, 047201 (2014)
22. D. Xiao, V. Tiberkevich, Y. H. Liu, Y. W. Liu, S. M. Mohseni, S. Chung, M. Ahlberg, A. N. Slavin, J. Åkerman, Y. Zhou, Parametric autoexcitation of magnetic droplet soliton perimeter modes. *Phys. Rev. B* **95**, 024106 (2017).
23. P. Fischer, Frontiers in imaging magnetism with polarized x-rays. *Front. Phys.* **2**, 1 (2015).
24. A. H. Reid, X. Shen, P. Maldonado, T. Chase, E. Jal, P. W. Granitzka, K. Carva, R. K. Li, J. Li, L. Wu, T. Vecchione, T. Liu, Z. Chen, D. J. Higley, N. Hartmann, R. Coffee, J. Wu, G. L. Dakovski, W. F. Schlotter, H. Ohldag, Y. K. Takahashi, V. Mehta, O. Hellwig, A. Fry, Y. Zhu, J. Cao, E. E. Fullerton, J. Stöhr, P. M. Oppeneer, X. J. Wang, H. A. Dürr, Beyond a phenomenological description of magnetostriction. *Nat. Commun.* **9**, 388 (2018).
25. A. von Reppert, L. Willig, J.-E. Pudell, S. P. Zeuschner, G. Sellge, F. Ganss, O. Hellwig, J. A. Arregi, V. Uhlír, A. Crut, M. Bargheer, Spin stress contribution to the lattice dynamics of FePt. *Sci. Adv.* **6**: eaba1142 (2020).
26. T. Henighan, M. Trigo, S. Bonetti, P. Granitzka, Z. Chen, M. Jiang, R. Kukreja, D. Higley, A. Gray, A. H. Reid, E. Jal, M. Hoffmann, M. E. Kozina, S. Song, M. Chollet, D. Zhu, P. F. Xu, J. Jeong, K. Carva, P. Maldonado, P. M. Oppeneer, M. G. Samant, S. S. P. Parkin, D. Reis, H. A. Dürr, Generation Mechanism of THz Coherent Acoustic Phonons in Fe. *Phys. Rev. B* **93**, 220301(R) (2016).
27. C. Dornes, Y. Acremann, M. Savoini, M.,Kubli, M. J. Neugebauer, E. Abreu, L. Huber, G. Lantz, C. A. F. Vaz, H. Lemke, E. M. Bothschafter, M. Porer, V. Esposito, L. Rettig, M. Buzzi, A. Alberca, Y. M. Windsor, P. Beaud, U. Staub, D. Zhu, S. Song, J. M. Glownia, S. L. Johnson, The ultrafast Einstein – de Haas effect. *Nature* **565**, 209 (2019).

28. J. Becker, O. Mosendz, D. Weller, A. Kirilyuk, J. C. Maan, P. C. M. Christianen, Th. Rasing, A. Kimel, Laser induced spin precession in highly anisotropic granular $L1_0$ FePt. *Appl. Phys. Lett.* **104**, 152412 (2014).
29. L. Dreher, M. Weiler, M. Pernpeintner, H. Huebl, R. Gross, M. S. Brandt, S. T. B. Goennenwein, Surface acoustic wave driven ferromagnetic resonance in nickel thin films: Theory and experiment. *Phys. Rev. B* **86**, 134415 (2012).
30. U. Atxitia, D. Hinzke, O. Chubykalo-Fesenko, U. Nowak, H. Kachkachi, O. N. Mryasov, R. F. Evans, R. W. Chantrell, Multiscale modeling of magnetic materials: Temperature dependence of the exchange stiffness. *Phys. Rev. B* **82,** 134440 (2010).
31. M. Poluektov, O. Eriksson, G. Kreiss, Coupling atomistic and continuum modelling of magnetism. *Comput. Methods Appl. Mech. Engrg.* **329**, 219 (2018).
32. B. S. D. Ch. S.Varaprasad, J. Wang, T. Shiroyama, Y.K. Takahashi, K. Hono, Columnar structure in fept-c granular media for heat-assisted magnetic recording. *IEEE Trans. Magn.* **51**, 3200904 (2015).
33. M. Porro, L. Andricek, S. Aschauer, A. Castoldi, M. Donato, J. Engelke, et al. The MiniSDD-Based 1-Mpixel Camera of the DSSC Project for the European XFEL. *IEEE Trans. Nucl. Sci.* **69**, 1334 (2021).
34. P. W. Granitzka, E. Jal, L. Le Guyader, M. Savoini, D. J. Higley, T. Liu, Z. Chen, T. Chase, H. Ohldag, G. L. Dakovski, W. Schlotter, S. Carron, M. Hoffmann, P. Shafer, E. Arenholz, O. Hellwig, V. Mehta, Y. K. Takahashi, J. Wang, E. E. Fullerton, J. Stöhr, A. H. Reid, H. A. Dürr, Magnetic switching in granular FePt layers promoted by near-field laser enhancement, *Nano Lett.* **17**, 2426 (2017).
35. A. Vansteenkiste, J. Leliaert, M. Dvornik, M. Helsen, F. Garcia-Sanchez, B. Van Waeyenberge, The design and verification of MuMax3. *AIP Adv.* **4**, 107133 (2014).
36. A. G. Gurevich, G. A. Melkov, Magnetization Oscillations and Waves. CRC Press Inc. (1996).
37. P. Nieves, S. Arpan, S. H. Zhang, A. P. Kadzielawa, R. F. Zhang, D. Legut, MAELAS: Magneto-ELAStic properties calculation via computational high-throughput approach. *Comp. Phys. Comms.* **264**, 107964 (2021).
38. D. Fritsch, C. Ederer, First-principles calculation of magnetoelastic coefficients and magnetostriction in the spinel ferrites $CoFe_2O_4$ and $NiFe_2O_4$. *Phys. Rev. B* **86**, 014406 (2012).
39. R. Xu, T. C. Chiang, Determination of phonon dispersion relations by X-ray thermal diffuse scattering. *Z. Kristallogr.* **220**, 1009 (2005).
40. A. Scherz, W. F. Schlotter, K. Chen, R. Rick, J. Stöhr, J. Lüning, I. McNulty, Ch. Günther, F. Radu, W. Eberhardt, O. Hellwig, S. Eisebitt, Phase imaging of magnetic nanostructures using resonant soft x-ray holography. *Phys. Rev. B* **76**, 1 (2007).
41. B. L. Henke, E. M. Gullikson, J. C. Davis, X-ray interactions: photoabsorption, scattering, transmission, and reflection at E=50-30000 eV, Z=1-92. *Atomic Data and Nuclear Data Tables* **54**, 181 (1993).
42. T. Chase, M. Trigo, A. H. Reid, R. Li, T. Vecchione, X. Shen, S. Weathersby, R. Coffee, N. Hartmann, D. A. Reis, X. J. Wang, H. A. Dürr, Ultrafast electron diffraction from non-equilibrium phonons in femtosecond laser heated Au films, *Appl. Phys. Lett.* **108**, 041909 (2016).
43. P. Maldonado, T. Chase, A. H. Reid, X. Shen, R. K. Li, K. Carva, T. Payer, M. Horn von Hoegen, K. Sokolowski-Tinten, X. J. Wang, P. M. Oppeneer, H. A. Dürr, Tracking the ultrafast nonequilibrium energy flow between electronic and lattice degrees of freedom on crystalline nickel, *Phys. Rev. B* **101**, 100302(R) (2020).
44. Turek, I., Drchal, V., Kudrnovský, J., Sob, M., Weinberger, P. Electronic Structure of Disordered Alloys, Surfaces and Interfaces (Kluwer, Boston, 1997).

45. Vosko, S. H., Wilk, L., Nusair, M. Accurate spin-dependent electron liquid correlation energies for local spin density calculations: A critical analysis, *Can. J. Phys.* **58**, 1200 (1980).
46. K. Yamamoto, Y. Kubota, M. Suzuki, Y. Hirata, K. Carva, M. Berritta, K. Takubo, Y. Uemura, R. Fukaya, K. Tanaka, W. Nishimura, T. Ohkochi, T. Katayama, T. Togashi, K. Tamasaku, T. Yabashi, Y. Tanaka, T. Seki, K. Takanashi, P. M. Oppeneer, H. Wadati, Ultrafast demagnetization of Pt magnetic moment in L10-FePt probed by magnetic circular dichroism at a hard x-ray free electron laser. *New J. Phys.* **21**, 123010 (2019).
47. S. V. Halilov, H. Eschrig, A. Y. Perlov, P. M. Oppeneer, Adiabatic spin dynamics from spin-density-functional theory: Application to Fe, Co, and Ni. *Phys. Rev. B* **58**, 293 (1998).
48. A. I. Liechtenstein, M. I. Katsnelson, V. P. Antropov, V. A. Gubanov, Local spin density functional approach to the theory of exchange interactions in ferromagnetic metals and alloys. *J. Magn. Magn. Mater.* **67**, 65 (1987).
49. P. M. Oppeneer, Magneto-optical spectroscopy in the valence-band energy regime: relationship to the magnetocrystalline anisotropy. *J. Magn. Magn. Mater.* **188**, 275 (1998).
50. Khan, S. A., Blaha, P., Ebert, H., Minar, J., Sipr, O. Magnetocrystalline anisotropy of FePt: A detailed view. *Phys. Rev. B* **94**, 144436 (2016).
51. S. Akiyama, Y. Tsunoda, Magnon and magnetic structure of FePt alloy. *J. Magn. Magn. Mat.* **310**, 1844 (2007).

**Acknowledgments**
The authors acknowledge European XFEL in Schenefeld, Germany, for provision of X-ray free-electron laser beamtime at Scientific Instrument SCS and thank the instrument group and facility staff for their assistance.
D.T., X.W and H.A.D. acknowledge support from the Swedish Research Council (VR), Grants 2017-06711 and 2018-04918.
A.Y. acknowledges support from the Carl Trygger Foundation.
V.U., N.Z.H. and S.B. acknowledge support from the European Research Council, Starting Grant 715452 MAGNETIC-SPEED-LIMIT.
E. J. is grateful for the financial support received from the CNRS-MOMENTUM program.
K.C. acknowledges support from the Czech Science Foundation (Grant No. 19-13659S).
P.M.O. acknowledges support by the Swedish Research Council (VR). Part of the calculations were enabled by resources provided by the Swedish National Infrastructure for Computing (SNIC) at NSC Linköping, partially funded by VR through Grant Agreement No. 2018-05973.
S.M. acknowledges support from the Ministry of Science and Higher Education of the Russian Federation (agreement no. 15.СИН.21.0003).
D.P. acknowledges support from the Russian Foundation for Basic Research (Grant no. 20-02-00489).
Work at the SLAC MeV-UED is supported in part by the DOE BES SUF Division Accelerator & Detector R&D program, the LCLS Facility, and SLAC under contract Nos. DE-AC02-05-CH11231 and DE-AC02-76SF00515.

**Author contributions:**
XFEL measurements: D.T., A.Y., X.W., V.U., I.V., M.S., E.J., R.C., G.M., R.G., N.A., B.K., L.M., J.S., L.G., N.G. M.T., D.L., D.P., D.M., J.B., N.Z.H., E.E.F., S.E., S.M., A.S., S.B., H.A.D.
UED measurements: D.T., I.V., A.H.R., X.S., X.J.W., H.A.D.

Sample growth and characterization: J.W., Y.K.T.
Micromagnetic simulations: E.I.
Magneto-elastic calculations: A.Y.
*Ab-inito* calculations: P.M., Y.K., K.C., P.M.O.
Data analysis: D.T., A.Y., X.W. I.V., M.S., E.I, S.B., H.A.D.
Writing – original draft: D.T., A.Y., X.W., E.I, S.B., H.A.D.
Writing – review & editing: all authors

**Competing interests:** Authors declare that they have no competing interests.
**Data and materials availability:** All data are available in the main text or the supplementary materials. Raw data generated at the European XFEL large-scale facility are available at: DOI: 10.22003/XFEL.EU-DATA-002599-00.

**Figures and Tables**

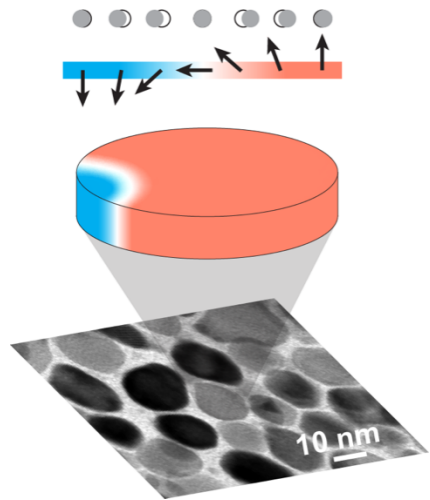

**Fig. 1. Schematic of FePt sample and magneto-elastic coupling.** The bottom panel shows a transmission electron microscopy image of FePt nanoparticles embedded in a C matrix (white). The grey-scale represents a spread in crystallographic alignment of the individual nanoparticles. The middle inset shows the average magnetization within one nanoparticle. The top inset displays the magnetization (colors and arrows) and atomic displacements (open and grey circles) across a magnetization texture.

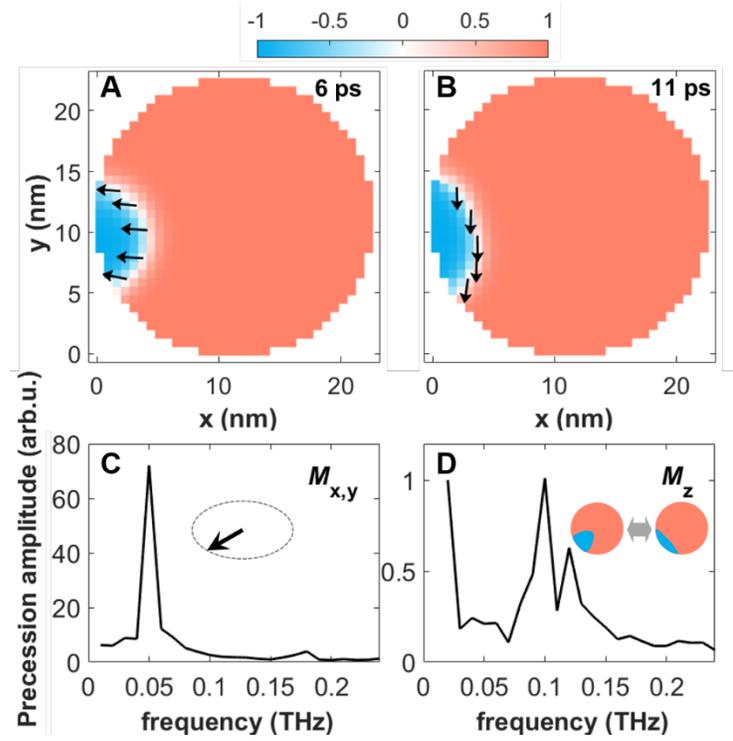

**Fig. 2. Magnetization dynamics of FePt spin-wave solitons from micromagnetic simulations. (A), (B)**, snapshots of the magnetization at 6 and 11 ps. **(C), (D)**, frequencies observed for the $M_{x,y}$ and $M_z$ magnetization components obtained via fast Fourier transforms of the full time-dependent simulations.

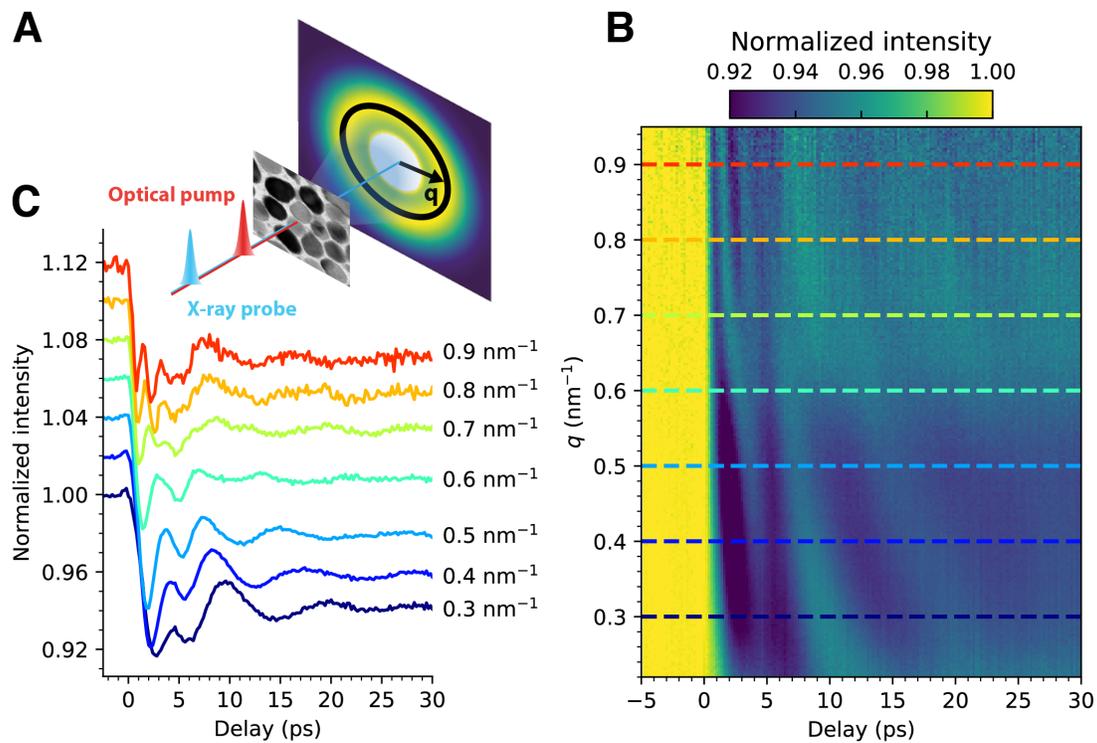

**Fig. 3. Time-domain measurements of FePt phonons.** (**A**), Optical pump – X-ray probe experimental geometry with the scattered wavevector, *q*, defined as indicated. (**B**), Time-delay map obtained by azimuthally averaging along the black circle in (**A**) and are normalized to the ground state (negative delay times). (**C**) Linecuts of the time-delay map at the indicated values of the wavenumber *q*, offset vertically for clarity.

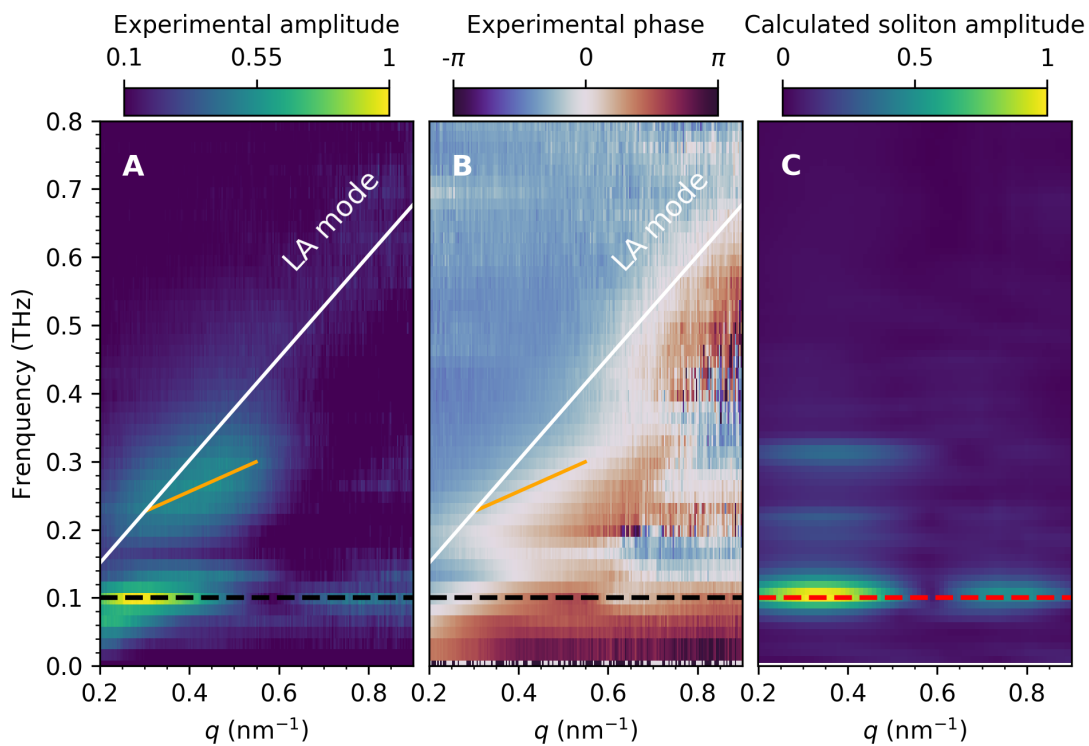

**Fig. 4. Characteristic frequencies of phonons generated by FePt spin-wave solitons.**
(**A**), amplitude and (**B**), phase in the frequency vs. wavevector representation of the time-domain data in Fig. 3B. The white lines show the calculated dispersion of the bulk FePt longitudinal acoustic (LA) phonon mode; the orange and dashed black lines mark the frequencies spin-wave soliton contributions, respectively. (**C**), the amplitude of calculated spin-wave soliton scattering contribution. The dashed red line shows the maximum of calculated spin-wave soliton precession contribution.

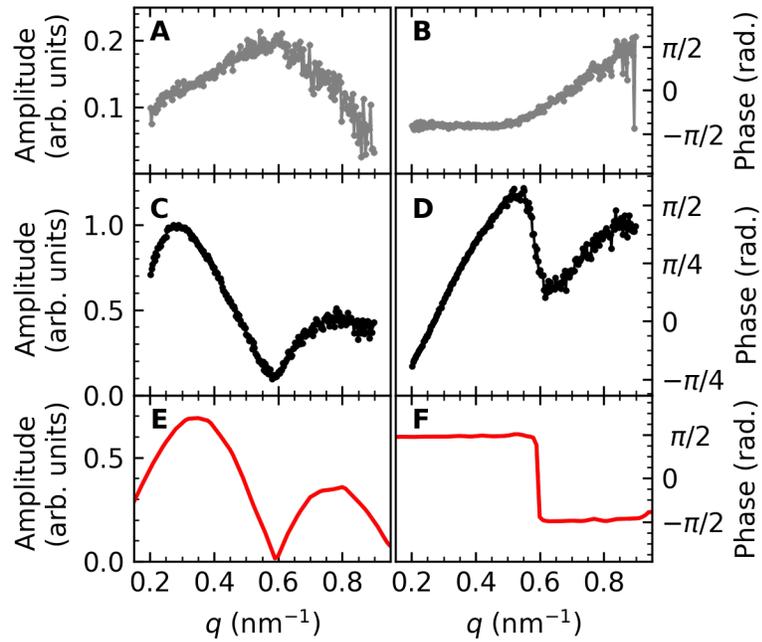

**Fig. 5. The amplitudes and phases of FePt lattice phonons at selected frequencies: (A), (B)** 0.50 THz, dominated by the LA phonon mode; **(C), (D)** 0.10 THz, where the maximum response of spin-wave soliton precession is observed; **(E), (F)** calculated amplitude and phase of the soliton contribution in the scattering at 0.10 THz.

**Table S1.** Elastic and magneto-elastic coupling parameters for the tetragonal $L1_0$ FePt phase.

| The elastic stiffness tensor constants (GPa) | | The magneto-elastic coupling constants (GPa) | |
|---|---|---|---|
| $C_{1111}$ | 254.8 | $b_{21}$ | 0.22 |
| $C_{1212}$ | 105.8 | $b_{22}$ | 0.08 |
| $C_{1313}$ | 117.7 | $b_3$ | 0.10 |
| $C_{1122}$ | 142.8 | $b'_3$ | 0.43 |
| $C_{1133}$ | 151.2 | $b_4$ | -0.05 |
| $C_{3333}$ | 318.8 | | |

**Supplementary Materials**

Supplementary Text
Figs. S1 to S7
Table S1
Movie S1

A model for FePt magneto-elastic coupling
This section describes in more detail the magneto-elastic coupling outlined in the methods section. The response of the FePt lattice structure to the presence of a spin-wave soliton in the nanoparticle was investigated using the magneto-elastic coupling formalism *(36)*, that describes the influence of magnetization dynamics on lattice atom displacements, **u**, in the material. In Cartesian coordinates, **x**, the magneto-elastic energy density for the tetragonal FePt lattice is given by the eq. (1) *(37,38)*:

$$\varepsilon_{mel}^{tet} = b_{11}(\varepsilon_{xx} + \varepsilon_{yy}) + b_{12}\varepsilon_{zz} + b_{21}\left(\alpha_z^2 - \frac{1}{3}\right)(\varepsilon_{xx} + \varepsilon_{yy}) + b_{22}\left(\alpha_z^2 - \frac{1}{3}\right)\varepsilon_{zz}$$
$$+ \frac{1}{2}b_3(\alpha_x^2 - \alpha_y^2)(\varepsilon_{xx} - \varepsilon_{yy}) + 2b'_3\alpha_x\alpha_y\varepsilon_{xy} + 2b_4(\alpha_x\alpha_z\varepsilon_{xz} + \alpha_y\alpha_z\varepsilon_{yz}), \quad (1)$$

where $\alpha_{x,y,z}$ are the direction cosines of the magnetization vector, **M**, $\varepsilon_{ij} = \frac{\partial u_i}{\partial x_j}$ are components of the strain tensor and $b_{11}, b_{12}, b_{21}, b_{22}, b_3, b'_3, b_4$ are tetragonal magneto-elastic coupling constants. Using the magneto-elastic energy density it is possible to calculate the magneto-elastic force:

$$\mathbf{f}_{mel} = \nabla\left(\frac{d\varepsilon_{mel}}{d\varepsilon_{ij}}\right) \quad (2)$$

For the tetragonal lattice eq. (2) leads to the following expression:

$$\mathbf{f}_{mel}^{tet} = \frac{1}{M_0^2}\begin{bmatrix} b_3 \frac{\partial M_x^2}{\partial x} \\ b_3 \frac{\partial M_y^2}{\partial y} \\ b_{22} \frac{\partial M_z^2}{\partial z} \end{bmatrix} + \frac{1}{2M_0^2}\begin{bmatrix} (b_3 + 2b_{21})\frac{\partial M_z^2}{\partial x} \\ (b_3 + 2b_{21})\frac{\partial M_z^2}{\partial y} \\ 0 \end{bmatrix} + \frac{1}{M_0^2}\begin{bmatrix} b_3' \frac{\partial}{\partial y}(M_x M_y) + b_4 \frac{\partial}{\partial z}(M_x M_z) \\ b_3' \frac{\partial}{\partial x}(M_y M_x) + b_4 \frac{\partial}{\partial z}(M_y M_z) \\ b_4 \frac{\partial}{\partial x}(M_z M_x) + b_4 \frac{\partial}{\partial y}(M_z M_y) \end{bmatrix} \quad (3)$$

In eq. (3) the first and the second terms are isotropic and the third term is the anisotropic component of the magneto-elastic force.

We note that all terms include spatial derivatives of the magnetization. This means that the force is nonzero only if the magnetization is not homogeneous over the nanoparticle volume. This implies that if the length of the magnetization vectors per unit volume does not change, the direction of the magnetization must change across the nanoparticle at any particular moment of time. As a consequence, homogeneous magnetic modes such as ferromagnetic resonance (FMR) cannot produce a considerable magneto-elastic coupling because the magnetic moments of all atoms in the nanoparticle precess as a whole with a similar amplitude and phase. Contrary to FMR, the spin-wave soliton has a well-defined profile that gives rise to strong magneto-elastic forces. These forces are especially large at the spin-wave soliton's perimeter. Here the $M_z$ component reverses sign while $M_{x,y}$ experiences a large-angle in-plane precession.

We calculate the details of the lattice dynamics in FePt nanoparticles by solving the equation of motion for the elastically strained solid with the magneto-elastic force resulting from the soliton magnetization dynamics as a driver for the atomic displacement:

$$\rho \frac{\partial^2 \mathbf{u}}{\partial t^2} + \frac{2\rho}{\tau}\frac{\partial \mathbf{u}}{\partial t} = \nabla \boldsymbol{\sigma} + \mathbf{f}_{mel} \quad (4)$$

where $\mathbf{u}$ is the displacement per unit volume, $\rho$ is the mass density, $\tau$ is a damping time constant, $\mathbf{f}_{mel}$ is the magneto-elastic force and $\nabla\boldsymbol{\sigma}$ is the elastic force per unit volume which defines the elastic properties of material and the phonon spectrum. It is given by the gradient of the three-dimensional stress tensor:

$$\sigma_{ij} = \sum_{k=1}^{3}\sum_{l=1}^{3} C_{ijkl}\varepsilon_{kl}, \quad (5)$$

where $C_{ijkl}$ are the components of the fourth-rank tensor of elastic stiffness parameters.

We calculate the magneto-elastic force $\mathbf{f}_{mel}$ using the expression eq. (3) with the results of micromagnetic simulation for an 8 nm high and 22.5 nm wide cylindrical nanoparticle as input. This is among the nanoparticle sizes commonly found in the sample. The equation of motion (4) was solved numerically in the 3-dimensional Cartesian grid using the standard second order "leapfrog" algorithm from the central differences. We used the same grid spacing as in the micromagnetic simulation: 0.7 nm in $x$, $y$ and 0.5 nm in $z$ directions. The time interval of simulation was 100 ps with the time step 50 fs. The initial lattice displacements were assumed to be zero, $\mathbf{u}_0 = 0$. For simplicity we have used the Dirichlet boundary condition $\mathbf{u}_{boundary} = 0$. The dissipation term with $\tau = 5$ ps was included in the eq. (4) to address, in the generalized form, the damping of magneto-elastically induced lattice vibrations via transmission through the nanoparticle boundary into the carbon matrix and other possible mechanisms.

Figure S4 illustrates the spin-wave soliton magnetization dynamics obtained from micromagnetic simulations described in the main paper (see methods section). The 22.5 nm nanoparticle displays a specific soliton mode with a frequency of 0.05 THz (Fig. S4F) for the in-plane magnetization components $M_x$ and $M_y$. This mode is best observed as a precession of the magnetization along the soliton's perimeter (marked in white in Figs. S4A-E). At the same time the dynamics of the out-of-plane $M_z$ component shows a lower-frequency peak near 0.10 THz that is related to the soliton breathing, i.e. periodic changes of its size and shape. In Figs. S4A-E, snapshots of the magnetization state demonstrate the spin-wave soliton precession and breathing dynamics. While the magnetization at the perimeter makes one full 360° rotation, the soliton shape changes about twice as fast. The soliton breathing can be clearly seen as the squeezed and elongated (blue) shapes in Figs. S4A-E (see also Movie S1).

Figure S5A shows the spectra of the magneto-elastic force calculated as the time-frequency Fourier transform of the $x, y, z$ force components in eq. (3) averaged over the particle's volume. The used FePt magneto-elastic coupling parameters are shown in Table S1 in the methods and materials section of the main paper. We see a strong peak at 0.10 THz and a weaker broader band between 0.2-0.3 THz. The former is caused mainly by the precession of the magnetization along the soliton's perimeter, while the latter corresponds to the magneto-elastic frequency mixing between the frequency-doubled in-plane magnetization precession and the spin-wave soliton breathing.

This assignment can be explained by considering a simplified model of the soliton magnetization precession and breathing modes. The $x, y,$ and $z$ components of magnetization near the soliton perimeter have the following form:

$$M_x(x,y) = M_0 \sin \theta_0(r(x,y)) \cos(\omega t)$$
$$M_y(x,y) = M_0 \sin \theta_0(r(x,y)) \sin(\omega t)$$
$$M_z(x,y) = M_0 \cos \theta_0(r(x,y)), \qquad (6)$$

where $\omega$ is the frequency of the soliton mode observed in $M_x$ and $M_y$ (Fig. S4F), that corresponds to the magnetization precession; while $\theta_0(r(x,y))$ is the spatial dependence of the polar magnetization angle for a spherical spin-wave soliton described in ref. *(7)*. $\theta_0$ is zero at the soliton perimeter but its detailed shape is only accessible to numerical calculations *(7)*. We will use $\theta_0$ here to parametrize the soliton shape. To address the spin-wave soliton breathing and it's coupling to phonons, the substitution can be made:

$$x \to x + \Delta x \cos\omega' t, \qquad (7)$$

where $\omega'$ is the soliton breathing frequency observed mainly in the $M_z$ component (Fig. S4F) and $\Delta x$ describes the breathing motion of the soliton perimeter which is "small" compared to the soliton size. Substituting the magnetization components calculated using eqs. (6) and (7) into eq. (3) the following approximate terms of the magneto-elastic force $x$ component can be derived:

$$f_{x1.1} = \frac{b_3}{2} \sin 2\theta_0 \frac{\partial \theta_0}{\partial r} \frac{(1 + \cos 2\omega t)(x + \Delta x \cos \omega' t)}{r}$$

$$f_{x1.2} = -\frac{(b_3 + 2b_{21})}{2} \sin 2\theta_0 \frac{\partial \theta_0}{\partial r} \frac{(x + \Delta x \cos \omega' t)}{r}$$

$$f_{x2} = \frac{b_3'}{2} \sin 2\theta_0 \frac{\partial \theta_0}{\partial r} \frac{\sin 2\omega t \, (y - \Delta x \cos \omega' t)}{r}$$

$$f_{x3} = b_4 \cos 2\theta_0 \frac{\partial \theta_0}{\partial z} \cos \omega t \tag{8}$$

In eqs. (8) we can notice the terms which contain the phases $2\omega t$ and $\omega' t$. Since the micromagnetic simulations of a 22.5 nm particle give $\omega \approx 0.05$ THz and $\omega' \approx 0.10$ THz (Fig. S4F), these terms apparently cause the peak in the force frequency spectrum at ~ 0.10 THz (Fig. S5A) as the second harmonic of $\omega$ and the fundamental $\omega'$ frequency. Eqs. (8) also contain cross-terms with the product of $2\omega t$ and $\omega' t$, that will lead to the addition (and subtraction) of these frequencies together resulting in the frequency band between 0.2-0.3 THz (and very low frequencies contributions that are not resolved experimentally). Thus, two frequency bands in the magneto-elastic force in Fig. S5A can be explained by the terms containing the doubled spin-wave soliton precession and the breathing frequency with $2\omega \approx \omega' \approx 0.10$ THz as well as the cross-terms between them. In other words, the interference between the magneto-elastic interactions caused by the soliton magnetization precession near the soliton perimeter and the soliton breathing produces two bands which differ approximately by a factor of two in frequency.

Furthermore, $f_{x1.1}$ in eq. (8) contain a time-independent term causing a constant force component near the domain wall, which is indeed observed in the full simulation. Finally, the term $f_{x3}$ may produce a linear frequency band near $\omega \approx 0.05$ THz. However, it is not observed in the experimental result because the soliton polar angle $\theta_0$ at the perimeter likely does not change along the $z$ direction, so that the derivative $\frac{\partial \theta_0}{\partial z}$ is close to zero. Also, the magneto-elastic constant $b_4$ responsible for this coupling is the smallest of all.

Figure S5B shows the calculated frequency spectrum of in-plane atomic displacements averaged over the particle volume. It shows a distinct peak at 0.10 THz and a weaker, broader band between 0.2-0.3 THz. The former corresponds well to the peak of the magneto-elastic force at the same frequency and is caused by the forced harmonic oscillations of the atoms with this frequency. The band at 0.2-0.3 THz overlaps with the band of the magneto-elastic force at the same frequency range and thus it is of induced nature as well. On the whole, the dominating lattice vibration bands at 0.10 and 0.2-0.3 THz, calculated for the 22.5 nm particle, are in good agreement with the soliton scattering contribution in experimental result in Fig. 4A of the main paper.

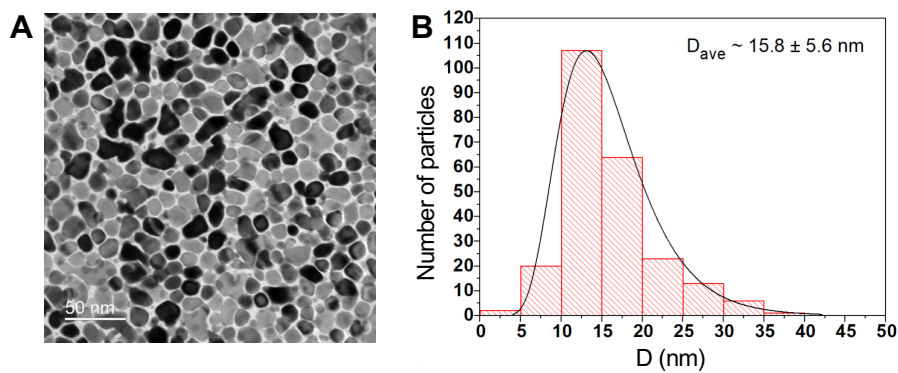

**Fig. S1.**
Nanoparticle size distribution of FePt samples. (**A**), transmission electron microscopy (TEM) image of FePt nanoparticles embedded in a C matrix (white). The different grey-scales visible in the FePt nanoparticles is caused by a spread in crystallographic alignment of the individual nanoparticles. (**B**), distribution of nanoparticle sizes, D. The average particle size evaluated from the TEM image is $D_{ave} = 15.8 \pm 5.6$ nm. The grain size distribution and average grain size were extracted from plane-view bright-field TEM images. More than 1,500 grains were randomly selected for the statistical analysis. The selected grains were covered with masks in the bright-field TEM image and the diameter information was acquired through analysis with the Gatan Microscopy Suite software. The collected data were fitted by a lognormal function to determine the average grain size and its standard deviation (error) to $D_{ave} = 15.8 \pm 5.6$ nm.

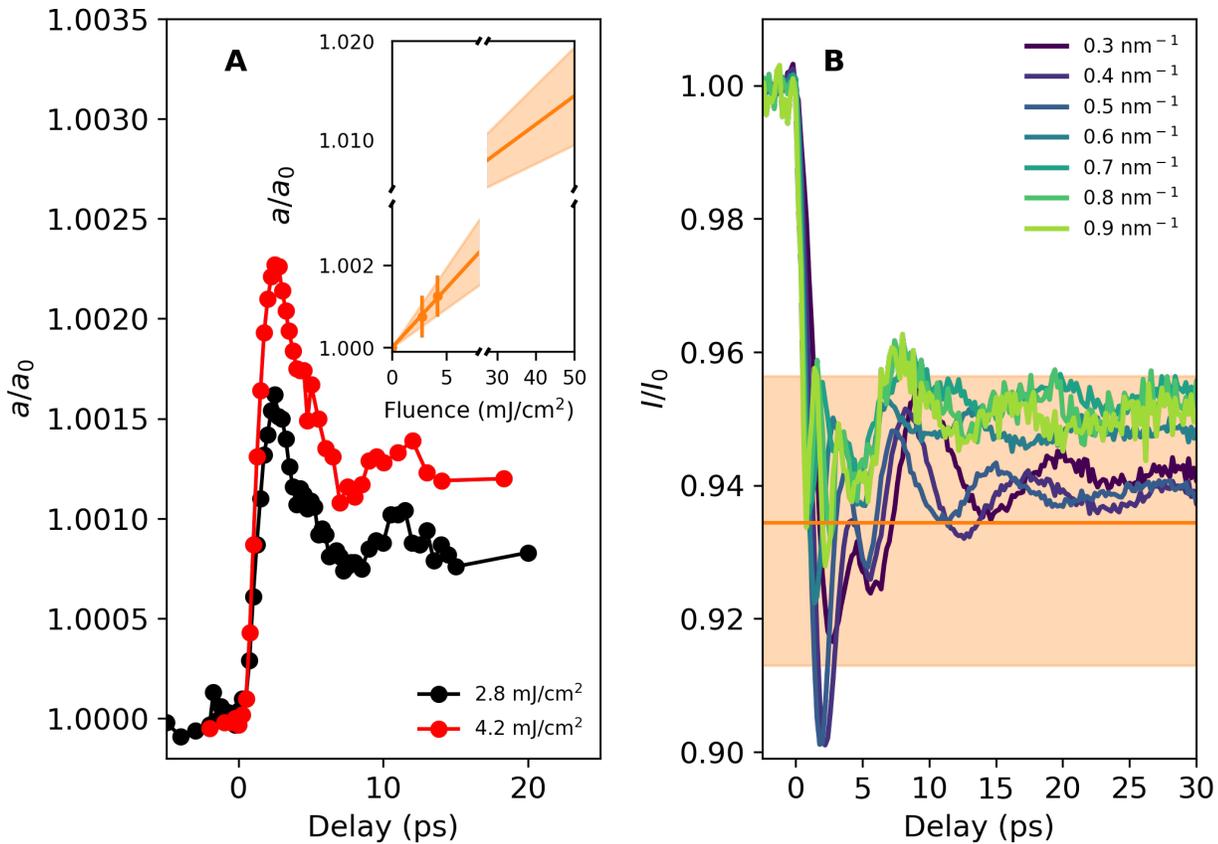

**Fig. S2.**
Effect of the lattice expansion on X-ray scattering contrast (**A**), Lattice expansion vs. pump-probe time delay for $L1_0$ FePt nanoparticles. Ultrafast electron diffraction (UED) was used to determine the changes of the lattice spacings parallel to the Fe and Pt layers (a-direction) of the $L1_0$ FePt lattice following fs laser excitation with 2.8 mJ/cm² (black symbols) and 4.2 mJ/cm² (red symbols), the inset shows that linear extrapolation to the fluence used in the main paper results in changes of $\Delta a/a_0 = +1.4\% \pm 0.5\%$ as obtained for delay times above ~5 ps where the values for $a/a_0$ are approximately constant. (**B**), scattering-intensity-drop calculated for the 1.4% change in lattice constant (orange line) overlaid on top of the measured X-ray scattering signal. The curves shown are the same as in Fig. 3 but this time they have not been shifted along the y-axis. The experimental results are within the extrapolation error (shaded in orange).

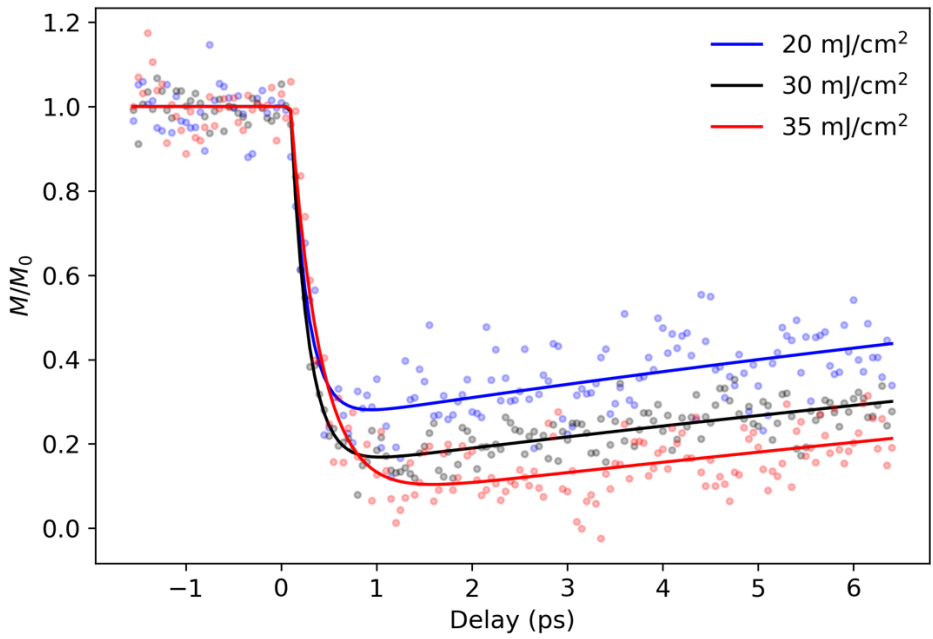

**Fig. S3.**
Ultrafast demagnetization of FePt for the indicated pump fluences. Following refs. *(23,33)* circularly polarized X-rays with energy tuned to the Fe $2p – 3d$ core-valence resonance (708 eV) were scattered from the FePt sample in the geometry shown in Fig. 3A. The sample magnetization was oriented (anti)parallel to the X-ray incidence direction by applying a magnetic field of $\pm 350$mT. The difference in the scattering yield, integrated over all accessible wavevectors, is a measure of the sample magnetization, *M*, and is plotted relative to the ground-state magnetization, $M_0$, measured before arrival of the pump pulse. The measured data (symbols) are fitted with an exponential decay followed by an exponential recovery (lines) of the form, $M/M_0 = 1 + A[exp(-\frac{t-t_0}{\tau_{rm}}) - exp(-\frac{t-t_0}{\tau_{dm}})]$ *(23,33)*. The fit parameters are compiled in Table S1.

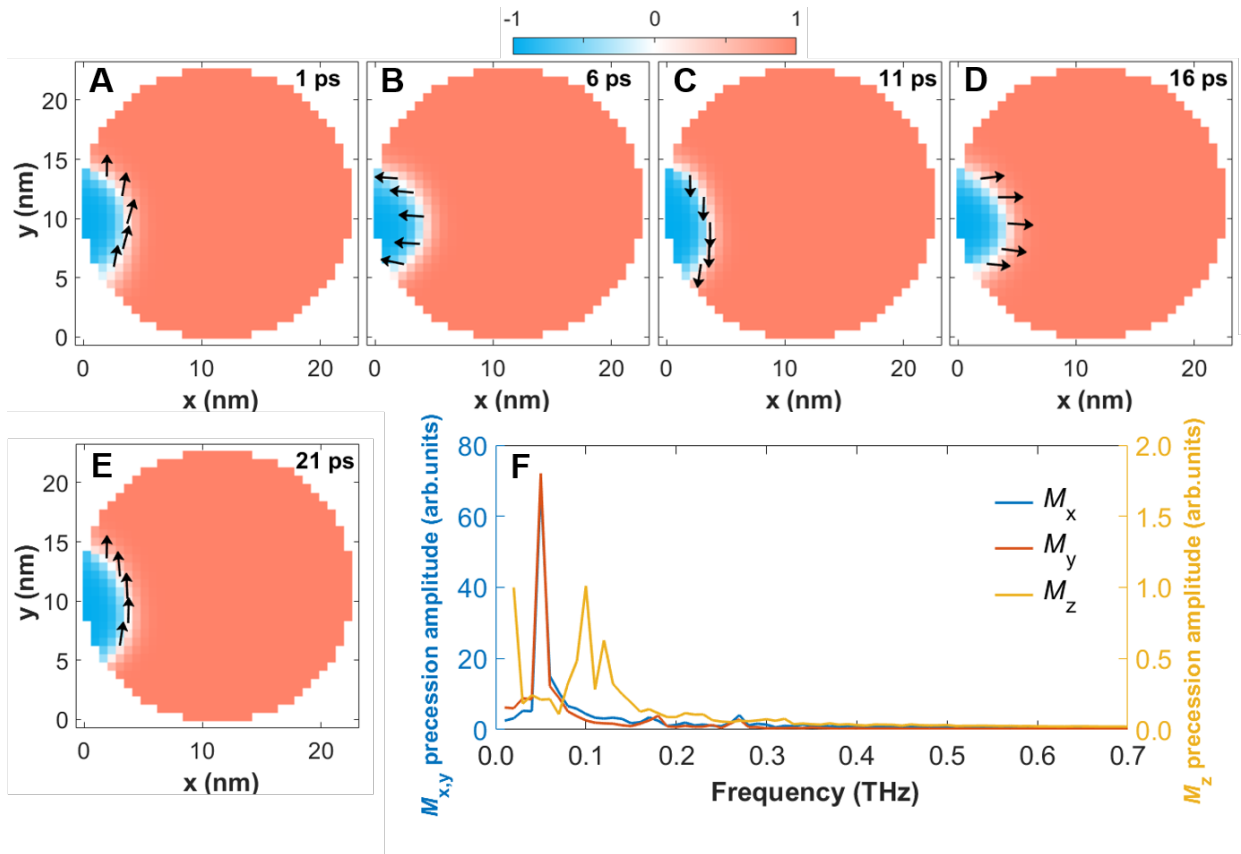

**Fig S4.**
(**A-E**), Snapshots of the micromagnetic simulation showing the correspondence between the two phases of soliton distortions (breathing and motion) and perimeter precession. The colour scale denotes the out-of-plane $M_z$ component. The black arrows show the magnetization in the perimeter that is predominantly oriented in-plane. (**F**), The Fourier transform amplitudes for the $M_x$, $M_y$ and $M_z$ magnetization components. The frequency of the $M_z$ precession corresponds to the spin-wave soliton breathing and motion.

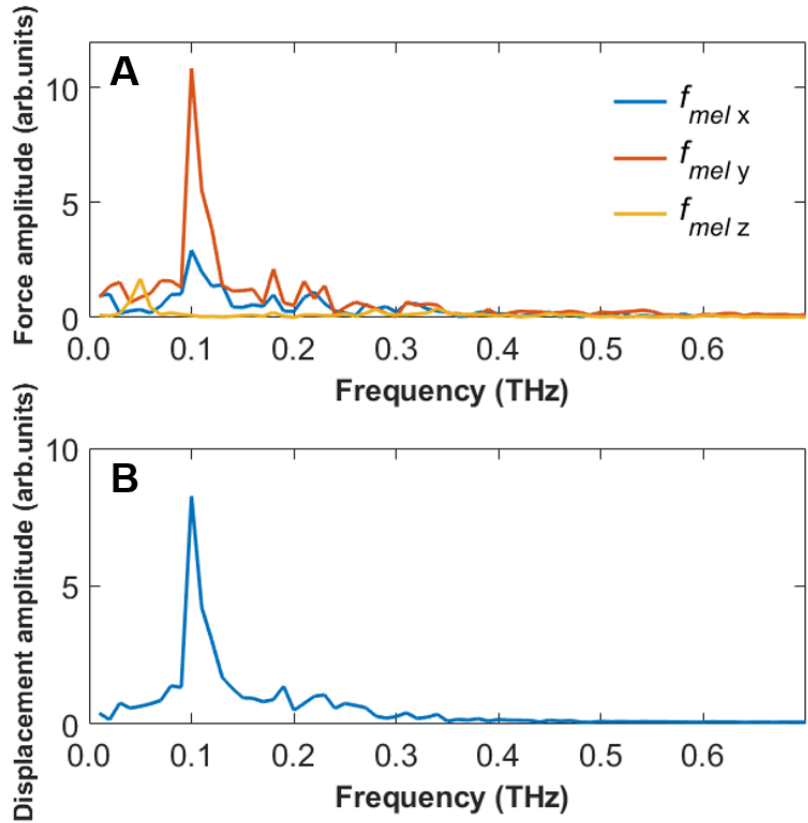

**Fig. S5.**

Results of the numerical magneto-elastic calculations: amplitudes of the Fourier transform of (**A**), magneto-elastic force and (**B**), atomic displacement. Both quantities show two dominating frequency bands at 0.10 and 0.2-0.3 THz.

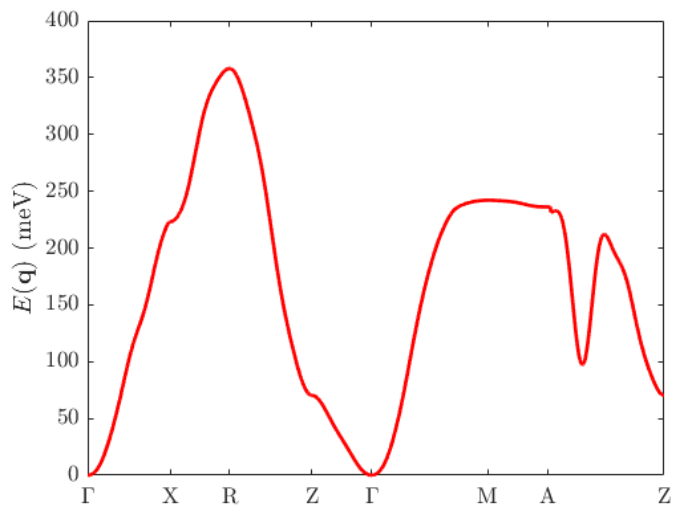

**Fig. S6.**
*Ab initio* calculated spin-wave dispersion of bulk FePt in the simple tetragonal Brillouin zone with corresponding high-symmetry points as indicated.

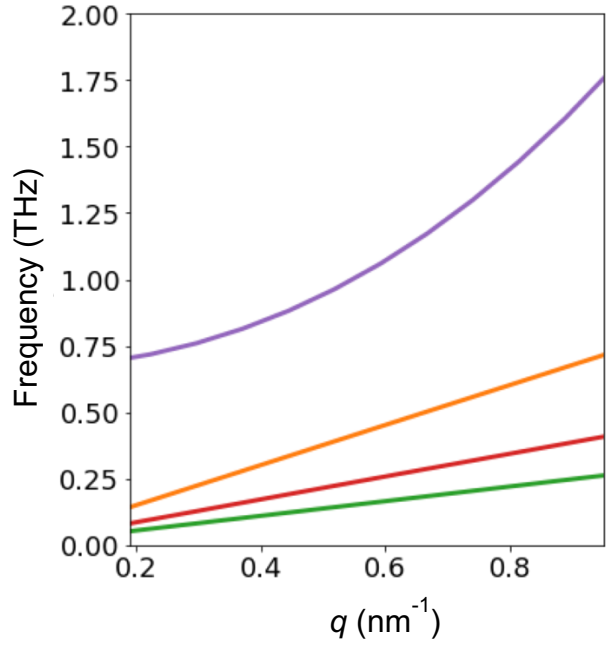

**Fig. S7.**
Calculations of magnon (purple line) and phonon (orange: longitudinal acoustic, LA, mode, red, green transverse acoustic, TA, modes) frequencies vs. wavevector, q.

**Table S1.**
The fitted parameters of the ultrafast magnetization dynamics in FePt of the form, $M/M_0 = 1 + A[exp(-\frac{t-t_0}{\tau_{rm}}) - exp(-\frac{t-t_0}{\tau_{dm}})]$ obtained from Fig. S3.

| Fluence (mJ/cm$^2$) | A | $\tau_{dm}$ (ps) | $\tau_{rm}$ (ps) |
|---|---|---|---|
| 20 | 0.754±0.017 | 0.176±0.021 | 21.4±2.9 |
| 30 | 0.863±0.009 | 0.184±0.010 | 29.9±2.6 |
| 35 | 0.943±0.018 | 0.314±0.021 | 34.5±5.7 |

**Other Supplementary Materials for this manuscript include the following:**
   Movie S1

**Movie S1.**

Movie of the temporal evolution of the magnetization of the FePt nanoparticle shown in Figs. 2A, B.